\documentclass{article}

\usepackage{arxiv}

\usepackage[utf8]{inputenc} 
\usepackage[T1]{fontenc}    
\usepackage{hyperref}       
\usepackage{url}            
\usepackage{booktabs}       
\usepackage{amsfonts}       
\usepackage{nicefrac}       
\usepackage{microtype}      
\usepackage{lipsum}		
\usepackage{graphicx,amssymb,amsmath,color}
\usepackage{doi}
\usepackage{ulem}
\usepackage{lineno}
\usepackage{array}
\newcolumntype{P}[1]{>{\centering\arraybackslash}p{#1}}
\usepackage{multirow}
\usepackage{enumitem}
\usepackage{tcolorbox}
\usepackage{xcolor}
\definecolor{ORNLgreen}{rgb}{0,.47,.2}
\definecolor{ORNLblue}{rgb}{0.34,.8,.6}

\usepackage{authblk}
\newcommand{\fig}{Fig.~}

\newcommand{\tbl}{Table~}

\newcommand{\sectn}{Section~}

\newcommand{\perspective}{Perspective~}

\usepackage{xr}
\makeatletter

\newcommand*{\addFileDependency}[1]{
\typeout{(#1)}
\@addtofilelist{#1}
\IfFileExists{#1}{}{\typeout{No file #1.}}
}\makeatother

\definecolor{blue}{rgb}{0, 0.5, 0.5}
\definecolor{blue2}{rgb}{0.1216, 0.4667, 0.7059}
\definecolor{red}{rgb}{0.8500, 0.1250, 0.0480} 
\definecolor{red2}{rgb}{0.8392, 0.1529, 0.1569} 
\definecolor{orange2}{rgb}{1.0, 0.498, 0.0549} 
\definecolor{yellow2}{rgb}{0.9290, 0.6940, 0.1250}
\definecolor{purple}{rgb}{0.4940, 0.1840, 0.5560}
\definecolor{purple2}{rgb}{0.5804, 0.4039, 0.7412}
\definecolor{green}{rgb}{0.4660, 0.6740, 0.1880}
\definecolor{green2}{rgb}{0.1725, 0.6275, 0.1725}
\definecolor{ltblue2}{rgb}{0.0902, 0.7451, 0.8118}
\definecolor{dkred2}{rgb}{0.6350, 0.0780, 0.1840}
\definecolor{gray2}{rgb}{0.22, 0.22, 0.3}
\definecolor{gray3}{rgb}{0.5, 0.5, 0.5}

\title{Exploring the use of quantum computing for facilitating spatially and temporally resolved models of a biological cell}

\author[a,$\dagger$]{Muralikrishnan Gopalakrishnan Meena}
\author[b]{Dileep Kishore}
\author[b]{Jerry M. Parks}
\author[c]{Luke Bertels}
\author[a]{Dilipkumar N. Asthagiri}
\author[d]{Travis Humble}
\author[a]{Thomas L. Beck}
\author[b,$\dagger$]{Mitchel J. Doktycz}

\affil[a]{National Center for Computational Sciences, Oak Ridge National Laboratory, Oak Ridge, TN 37831, USA}
\affil[b]{Biosciences Division, Oak Ridge National Laboratory, Oak Ridge, TN 37831, USA}
\affil[c]{Computational Sciences and Engineering Division, Oak Ridge National Laboratory, Oak Ridge, TN 37831, USA}
\affil[d]{Quantum Science Center, Oak Ridge National Laboratory, Oak Ridge, TN 37831, USA}
\affil[$\dagger$]{To whom correspondence should be addressed: \href{mailto:gopalakrishm@ornl.gov}{gopalakrishm@ornl.gov}, \href{mailto:doktyczmj@ornl.gov}{doktyczmj@ornl.gov}}

\date{}

\renewcommand{\headeright}{~}
\renewcommand{\undertitle}{~}
\renewcommand{\shorttitle}{Quantum computing for spatially and temporally resolved models of a biological cell}

\begin{document}

\maketitle

\tableofcontents

\begin{abstract}
\label{sec:abstract}
Whole-cell simulation, modeling all of a cell's functional systems over its life cycle, is an outstanding challenge in computational biology. Even the simplest living cell contains thousands of interacting proteins and metabolites (on the order of trillions of atoms) whose full functional dynamics spans roughly five orders of magnitude in space (nm to $\mu$m) and nearly nineteen in time (fs to hours). Further, many of the governing physical and chemical properties remain incompletely characterized. Simulating such complex systems at fully atomistic resolution over a full cell cycle is computationally intractable on classical architectures, raising a central question: Can quantum computing offer a viable path to whole-cell simulations that integrate molecular- and systems-level complexity? This \perspective examines the potential of quantum computing across three hierarchical scales: atomistic-molecular modeling, metabolic and regulatory networks, and whole-cell spatial modeling. We present a complexity analysis comparing classical and quantum algorithms for representative biological problems, identifying regimes of substantial theoretical speedup under specified algorithmic assumptions. We highlight algorithmic developments designed to leverage both near-term exploratory and fault-tolerant quantum architectures, and discuss practical bottlenecks: data encoding overhead, system conditioning, measurement constraints, and hybrid quantum-HPC integration. Together, these results outline a roadmap for quantum-accelerated whole-cell modeling and the biological insights such multiscale frameworks may eventually enable.

\footnote{\textbf{Notice:} This manuscript has been authored by UT-Battelle, LLC under contract DE-AC05-00OR22725 with the US Department of Energy (DOE). The US government retains and the publisher, by accepting the article for publication, acknowledges that the US government retains a nonexclusive, paid-up, irrevocable, worldwide license to publish or reproduce the published form of this manuscript, or allow others to do so, for US government purposes. DOE will provide public access to these results of federally sponsored research in accordance with the DOE Public Access Plan (\href{http://energy.gov/downloads/doe-public-access-plan}{http://energy.gov/downloads/doe-public-access-plan}).}
\end{abstract}

\section*{Plain language summary}

Living cells are among the most complex systems in nature, with trillions of interacting molecules spanning across scales of size and time so vast that simulating even a single simple cell over its entire life cycle from first principles is beyond the capability of today's most powerful supercomputers. Quantum computers process information in a fundamentally different manner and may eventually overcome some of the these barriers. We survey how quantum algorithms could potentially accelerate the three major layers of whole-cell modeling --- the chemistry of individual molecules, the networks of reactions amongst the molecules, and the movement of molecules through the cellular space. We identify where the largest gains are theoretically possible with some candidate problems for each of the three layers. By mapping both the potential opportunities as well as practical challenges, this work offers a roadmap toward simulating a living cell in full molecular detail, a long-standing goal that could transform how we understand health, disease, and the basic machinery of life.

\section{Introduction}\label{sec:intro}

The advent of quantum-based computational approaches offers new promise in addressing currently intractable problems in biology \cite{emani2021quantum,marx2021biology,marchetti2022quantum,blunt2022perspective,cordier2022biology,mcweeney2023quantum,pal2024quantum,pyrkov2024complexity}. One of the most formidable challenges in biology is the development of accurate and comprehensive whole-cell simulations \cite{singla2021community,stumpf2021statistical}. Even so-called “simple” cells contain trillions of interacting biomolecular components operating across vast spatial and temporal scales, while the physical and chemical properties governing many of these interactions remain incompletely characterized. The collective behavior of these components gives rise to complex molecular systems and emergent properties at higher levels of organization, phenomena that continue to challenge traditional modeling and simulation techniques.

Simulating such compositionally and dynamically complex systems over biologically relevant time scales remains computationally impractical using conventional computing architectures. As Netz and Eaton \cite{netz2021estimating} have described, realizing cellular descriptions with molecular and atomic resolution is simply intractable. Nonetheless, a diverse set of modeling strategies has been developed, providing valuable mechanistic insights and enabling hypothesis generation and testing. These approaches can be broadly categorized as models that emphasize 1) molecular level interactions, 2) biochemical networks, or 3) whole-cell simulations \cite{georgouli2023multi}. Each class relies on distinct experimental data and computational formalisms, yet all remain fundamentally constrained by the performance limits of conventional computing platforms. This challenge raises a central question: Can quantum-based computational platforms overcome these limitations, and do they offer a viable path toward whole-cell simulations that accurately integrate molecular- and systems-level complexity?

\begin{figure}[ht!]
\centering
\includegraphics[width=1\textwidth]{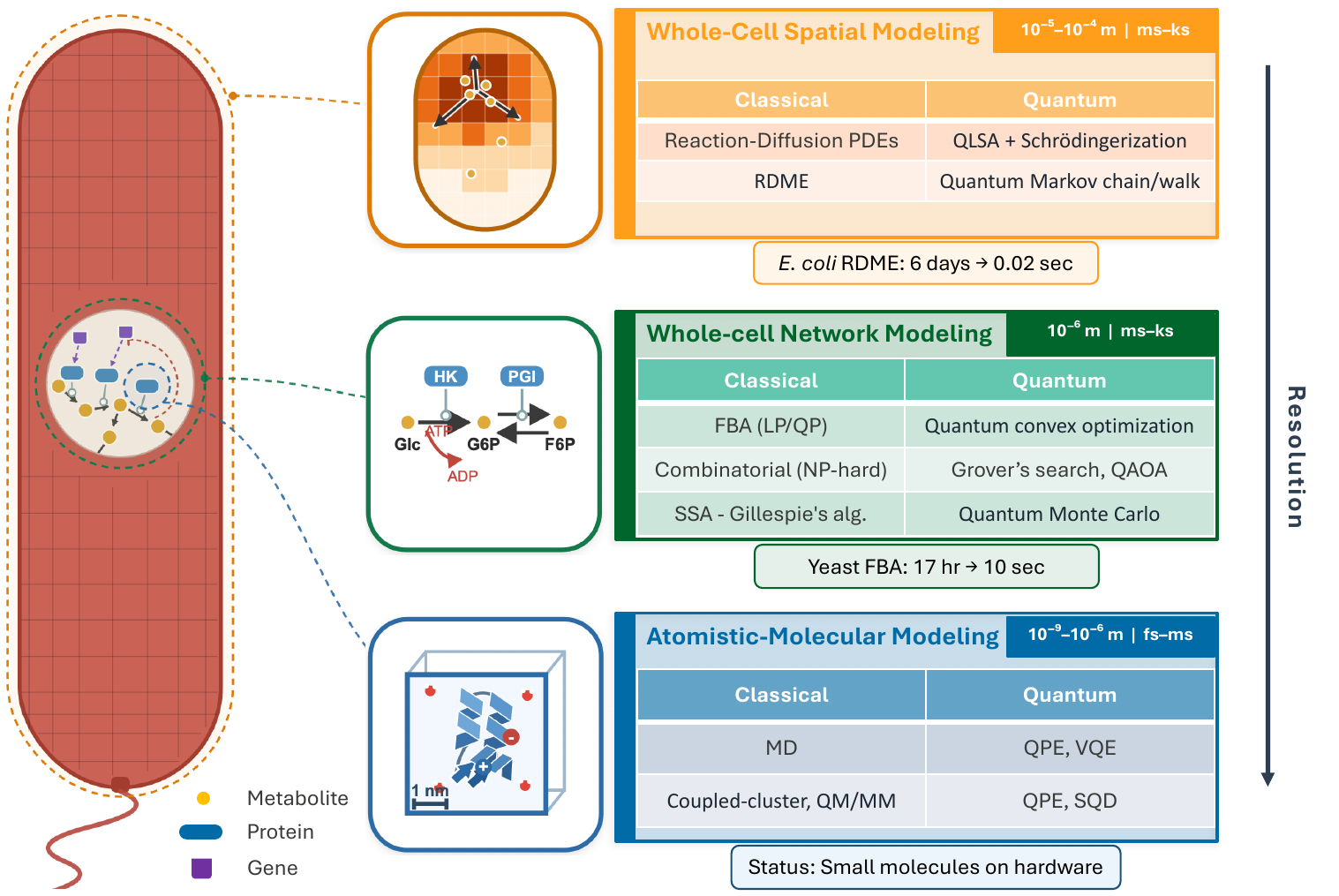}
\caption{Classical and quantum computational methods across scales of whole-cell modeling. A bacterial cell, shown as an intracellular network of metabolites, proteins, and genes, is resolved at three successively finer scales (arrow, increasing spatial resolution), with color-matched lenses linking each regime to the cellular feature it captures: the whole cell (spatial; orange), a subnetwork (network; green), and a single macromolecule (atomistic–molecular; blue). For each regime, the table compares classical methods with their emerging quantum counterparts, listing the relevant length and time scales and an illustrative reported speed-up. Icons depict a representative system at each scale. }\label{fig:outline}
\end{figure}

In this \perspective, we examine the potential of quantum-based computational platforms to advance simulations across molecular, systems, and cellular scales. These three hierarchical scales with the order of spatial and temporal scales, classical approaches, and potential quantum alternatives are outlined in \fig\ref{fig:outline}. We evaluate the extent to which existing quantum algorithms may offer advantages over classical numerical methods when applied to established modeling strategies, and we highlight emerging algorithmic developments designed to explicitly leverage quantum computing architectures. We further discuss the technical steps required for the practical deployment of quantum computing systems in the context of biological modeling. Finally, we consider the transformative potential of whole-cell simulations that integrate biochemical reaction networks with molecular-level precision, and the new biological insights that may emerge from such multiscale, quantum-enabled frameworks.

\section{Modeling techniques}\label{sec:modeling}

Quantum computing offers a range of algorithmic strategies for accelerating the classical modeling techniques used at each of the three scales of whole-cell modeling introduced above. To assess where these strategies may offer genuine advantages, we perform a systematic complexity analysis, comparing the computational scaling of established classical methods against candidate quantum approaches for representative problems at each scale (\tbl\ref{tab:quantum_classical_comparison}). In the following subsections, for each scale, we describe the relevant classical methods and their computational bottlenecks, survey the quantum algorithms proposed to address them, and highlight a state-of-the-art hardware demonstration at the end of each subsection. While we assess the theoretical scaling of the various quantum algorithms in this section, their practical implementation is often hindered by algorithmic and hardware challenges, including data encoding into the quantum system, measurement of the outputs (or readout constraints), noise, and error correction requirements. We discuss the practical challenges for implementing these approaches in \sectn\ref{sec:discussion}.

\begin{table}[t]
\centering
\caption{Classical and quantum methods across biological scales for whole-cell modeling. fs = femtoseconds, $\mu$s = microseconds, ms = milliseconds, s = seconds, ks = kiloseconds. $\kappa$ = condition number, $R$ = nonlinearity/dissipation ratio, $N$ = system dimension.}
\begin{tabular}{p{2.5cm}p{1.8cm}p{3.5cm}p{4cm}p{2.8cm}}
\hline
\textbf{Scale} & \textbf{Timescale} & \textbf{Classical Method} & \textbf{Quantum Approach} & \textbf{Status} \\
\hline
\multicolumn{5}{l}{\textit{Atomistic (10$^{-9}$-10$^{-6}$ m)}} \\
\hline
Macromolecular structure prediction & $\mu$s-ms & Classical MD, force fields~\cite{stevens2023molecular} & VQE, CVaR-VQE on lattice models~\cite{tilly2022variational} & 7-10 amino acids on quantum hardware \cite{robert2021resource} \\
Quantum chemistry & fs-$\mu$s & DFT, coupled-cluster & VQE, QAOA, SQD~\cite{vqe,farhi2014quantumqaoa,shajan2026molecular} & Small molecules (VQE) to 300-atom protein via embedding + SQD~\cite{vqe,shajan2026molecular} \\
Protein dynamics & $\mu$s-s & Gaussian network models & Hamiltonian simulation~\cite{low2017optimal,low2019hamiltonian} & Theoretical framework \\
\hline
\multicolumn{5}{l}{\textit{Molecular Networks (10$^{-6}$ m)}} \\
\hline
Gene regulatory networks & ms-ks & Boolean networks, probabilistic BN~\cite{boyd2004convex,nocedal2006numerical} & qscGRN, Grover-inspired attractor search~\cite{roman2023quantum} & Single-cell RNA-seq data, 5-node networks~\cite{roman2023quantum} \\
Pathway models & s-ks & FBA, SBML models~\cite{orth2010flux,palsson2015systems} & Grover search for SAT~\cite{grover1996fast,brisebois2025identifying} & Protein regulatory logic on hardware~\cite{brisebois2025identifying} \\
\hline
\multicolumn{5}{l}{\textit{Cellular - Deterministic (10$^{-5}$-10$^{-4}$ m)}} \\
\hline
ODE systems & s-ks & RK methods, implicit solvers~\cite{trefethen2022numerical} & HHL, QLSA variants~\cite{harrow2009quantum,morales2024quantum} & Small-scale demos; requires $\kappa < N$~\cite{yalovetzky2024solving,zheng2024early} \\
Reaction-diffusion PDEs & s-ks & FEM, FDM, spectral methods & HHL + Carleman linearization~\cite{liu2021efficient,lockwood2025quantum} & Requires strong dissipation ($R \ll 1$)~\cite{liu2021efficient} \\
Navier-Stokes & $\mu$s-ks & FVM, FDM & Carleman + QLSA~\cite{gaitan2020finding,meena2024towards} & 1D-2D incompressible flow demos~\cite{gopalakrishnan2024solving,song2025incompressible} \\
\hline
\multicolumn{5}{l}{\textit{Cellular - Stochastic (10$^{-5}$-10$^{-4}$ m)}} \\
\hline
Non-spatial & ms-ks & Gillespie SSA, tau-leaping~\cite{gillespie1977exact,gillespie2007stochastic} & QTT (quantum-inspired classical)~\cite{kazeev2014direct} & Demos of the algorithms on non-bio apps \\
Spatial & ms-ks & RDME, particle-based~\cite{ghaemi2020silico,johnson2021quantifying} & Quantum Monte Carlo, SPDE with Schrödingerization, Quantum Markov Chain, Quantum walk-based algorithms~\cite{an2021quantum,jin2025quantum,szegedy2004quantum,venegas2012quantum} & Demos of the algorithms on non-bio apps~\cite{an2021quantum,layden2023quantum} \\
\hline
\end{tabular}
\small
\label{tab:quantum_classical_comparison}
\end{table}

\subsection{Molecular simulations}\label{subsec:model-atomic}
Atomistic and molecular modeling seeks to resolve the chemistry and physics that ultimately set the effective parameters of higher-scale cellular models. A fully atomistic description of an entire cell, however, remains far beyond reach. Thus, whole-cell simulation efforts such as the 4D model of the genetically minimal bacterium JCVI-syn3A rely on coarse-grained or other reduced representations rather than atomistic detail \cite{thornburg2026bringing,stevens2023molecular}. The central question for this scale is therefore where quantum computing can either improve the accuracy of the chemical inputs to these models or supply information that is currently unavailable to them.

The conventional biomolecular simulation toolkit spans several mature methods. Electronic structure may be described with density functional theory (DFT), perturbation theory (e.g., MP2), coupled-cluster (CC), or complete-active-space approaches (CASSCF/CASPT2) for active sites, redox chemistry, and charge transfer. In addition, semiempirical and tight-binding methods such as DFTB are widely used because they are much less computationally expensive than DFT, albeit at some loss of accuracy. Classical MD with molecular mechanics (MM) force fields capture femtosecond-to-microsecond dynamics and is routinely combined with enhanced-sampling methods to simulate large-scale conformational changes and estimate thermodynamic quantities. Hybrid quantum mechanical/molecular mechanical (QM/MM) schemes embed electronic structure representations of enzyme active sites into MM representations of the rest of the protein and solvent environment \cite{senn2009qm}, thereby enabling simulations of enzymatic reactions. Initial structures of proteins are obtained from the Protein Data Bank \cite{berman2002protein,berman2003announcing}, but are increasingly complemented by AlphaFold2/3 \cite{jumper2021highly,abramson2024accurate} and related models. Generating simulation-ready systems requires careful preparation. Three main bottlenecks limit the predictive accuracy of this stack. 

First, \emph{electronic structure accuracy in chemically complex regimes}: Standard methods are often inaccurate for systems with substantial multi-reference character. DFT approximations exhibit functional-dependent errors for transition metals, spin-state ordering, redox energetics, and reaction barriers. Large-scale CC and CAS treatments scale prohibitively for realistic active sites, and CAS treatments require delicate active-space selection \cite{reiher2017elucidating}. These issues are particularly important for metalloenzymes, open-shell intermediates, and strongly correlated active sites, but they are not the primary limitation for all enzymes. Second, \emph{configurational sampling}: rare events dominate kinetics and free energies, and brute-force MD fails to converge barrier crossing and ligand binding on biological time scale. A central target is the free energy difference $\Delta F = -k_\mathrm{B}T\ln(Z_1/Z_0)$ between states, where $Z_0$ and $Z_1$ are the partition functions of two different states of the system \cite{netz2021estimating}. In practice, mature techniques such as umbrella sampling, metadynamics, free energy perturbation, thermodynamic integration, Markov state models, and related enhanced sampling methods are widely used. Third, \emph{environmental complexity}: macromolecular crowding, pH, ionic strength, dielectric heterogeneity, and protonation/tautomer-state uncertainty shift conformational ensembles and reaction energetics, and these uncertainties propagate upward as the effective rate constants, binding free energies, $\mathrm{p}K_\mathrm{a}$ values, and redox potentials consumed by cellular-scale models \cite{ellis2001macromolecular,thornburg2026bringing,ghaemi2020silico}.

Among the three modeling scales considered in this \perspective, atomistic-molecular simulation is where quantum computing is most developed, both algorithmically and in hardware demonstrations, and it admits two complementary roles: improving the \emph{accuracy} of chemical inputs (energies, forces, ground states) to existing classical workflows, and, in the longer term, providing \emph{first-principles information} that replaces empirically fitted parameters altogether. In the fault-tolerant limit, quantum simulation algorithms such as quantum phase estimation (QPE) can recover molecular energies from a Hamiltonian eigenvalue problem \(\hat{H}\lvert\psi\rangle=E\lvert\psi\rangle\) \cite{kitaev1995quantum}, provided that a sufficiently accurate approximate eigenstate can be prepared. Modern fault-tolerant Hamiltonian simulation based on qubitization and quantum signal processing (QSP) can reduce the asymptotic overhead for implementing \(e^{-i\hat{H}t}\)  once an appropriate block encoding or related oracle construction is available \cite{low2017optimal,low2019hamiltonian}. 

These approaches are most relevant for metalloenzyme energetics, spin-state ordering, and reaction barrier heights where classical DFT may be unreliable and high-level CC or large-CAS treatments can be intractable for realistic active sites. Nearer-term workflows instead emphasize embedding and fragmentation so that chemically decisive regions can be treated at higher electronic resolution. A recent protein-scale demonstration couples an embedding construction of reduced fragment Hamiltonians with sample-based quantum diagonalization (SQD) as a hardware subroutine, and reconstructs conformer energetics across a folded peptide system \cite{shajan2026molecular}. This embedding-plus-SQD approach treats the quantum device as a fragment eigensolver integrated with classical pre- and post-processing, rather than relying on deep phase-estimation circuits. This approach is also a natural way to supply improved initial states for downstream simulations. More broadly, such hybrid approaches are attractive because they target localized, chemically important subproblems while preserving compatibility with established classical simulation workflows. Table~\ref{tab:atomistic_quantum} organizes the principal quantum approaches by hardware era, from near-term variational methods to long-term fault-tolerant targets, together with their demonstrated status and projected resource requirements.

\begin{table}[t]
\centering
\caption{Quantum approaches for atomistic-molecular simulation, organized by hardware era. Resource targets are order-of-magnitude estimates from the cited demonstrations and surveys. NISQ = Noisy Intermediate-Scale Quantum device, FT = fault-tolerant.}
\label{tab:atomistic_quantum}
\begin{tabular}{p{2.0cm} p{3.0cm} p{4.0cm} p{2.6cm} p{2.4cm}}
\hline
\textbf{Era} & \textbf{Method} & \textbf{Status / demo} & \textbf{Resource target} & \textbf{Use case} \\
\hline
NISQ & VQE, CVaR-VQE, QAOA & Small molecules (H$_2$, LiH); 7--10 AA peptides & Tens of qubits & Small molecules; folding \\\\

Hardware demo & Embedding + SQD & Folded peptide \cite{shajan2026molecular,merz2026crossing} & Current devices + classical pre/post & Conformer energetics \\\\

Near-term FT & QPE, small-CAS & Theoretical: small chemistry \cite{alexeev2025perspective} & 25--100 logical qubits & Reactions, binding \\\\

Long-term FT & QPE + qubitization / QSP & Theoretical: nitrogenase \cite{reiher2017elucidating} & $\sim$2{,}000 logical qubits, $\sim$10$^{14}$ T-gates & Enzyme mechanisms \\
\hline
\end{tabular}
\end{table}

The trajectory across these eras points toward a clear long-term objective: progressively reduce the requirement for 
coarse-graining by making first-principles energetics available at scales that are currently inaccessible. Quantum advantage is expected to emerge most clearly for strongly correlated active spaces and high-accuracy energy differences that dominate downstream uncertainty, because fault-tolerant algorithms can, in principle, reach chemical accuracy with costs that scale more favorably than CC or large-CAS treatments as complexity increases. As quantum resources and algorithms mature, a credible end point is fully atomistic, chemically accurate cellular simulation in which rates and forces are derived from consistent electronic structure across many interacting subsystems, thereby replacing large fractions of empirically fitted effective parameters with transferable, first-principles descriptions of reactivity and intermolecular interactions. These developments can inform in a more natural way the quantitative modeling of the interaction among the components in the regulatory networks in cells at the next level (\fig\ref{fig:outline}) of spatial and temporal resolution. This capability is especially important for biochemical reactions involving bond breaking and formation, charge transfer, and changes in spin or protonation state. For such processes, density-functional approximations can exhibit delocalization and localization errors, fractional-electron and fractional-spin errors, and functional-dependent errors in reaction barriers and redox energetics. These limitations can propagate into the thermodynamic and kinetic parameters used by higher-level cellular models.

Accurate molecular thermodynamics and kinetics, including reaction free energies, activation barriers, equilibrium constants, and rate constants, provide essential inputs to the metabolic and regulatory models considered in Section 2.2. Quantum-derived calculations could therefore improve the parameterization of reaction networks where experimental measurements are unavailable, uncertain, or difficult to obtain under intracellular conditions. They could also generate high-fidelity, physically constrained reference data for AI/ML models that predict molecular properties or serve as surrogate models within multiscale simulations. In this way, improved electronic-structure calculations can inform the quantitative modeling of interactions among cellular components at the next level of spatial and temporal resolution (\fig\ref{fig:outline}).

\begin{tcolorbox}[colframe=black, colback=ORNLgreen!60, sharp corners=southwest, boxrule=1pt, width=\textwidth, left=2mm, right=2mm, top=1mm, bottom=1mm]
\label{box:practical_atomistic}
\textbf{State of the art --- protein electronic structure on quantum hardware.}
A 303-atom Trp-cage miniprotein, fragmented by wave-function embedding and solved with SQD on IBM Heron hardware (largest fragment: 33 orbitals), yields relative conformer energies benchmarked against CCSD~\cite{shajan2026molecular}. This pushes quantum electronic-structure calculation from small molecules toward protein scale on present-day devices.
\end{tcolorbox}

\subsection{Whole-cell metabolic and regulatory network modeling}\label{subsec:model-network}

Cellular function emerges from the coordinated activity of metabolic and regulatory networks comprising thousands of interacting components. Computational modeling of these networks spans a hierarchy of approaches, each trading off mechanistic detail against data requirements and computational tractability. At one extreme, constraint-based models capture steady-state flux distributions without requiring kinetic parameters; at the other, stochastic models resolve the discrete, probabilistic nature of individual molecular events. Between these lie kinetic models based on ordinary differential equations (ODEs), which capture temporal dynamics but demand extensive parameterization. Understanding the computational challenges and quantum opportunities at each level provides essential context for whole-cell modeling efforts.

At the genome scale, constraint-based models such as flux balance analysis (FBA) represent metabolism using a stoichiometric matrix ($S$) subject to steady-state mass balance and flux bounds ($S\mathbf{v} = 0$), yielding linear or quadratic optimization problems (LP/QP) typically solved via simplex or interior-point methods with polynomial complexity scaling as $\mathcal{O}(N^{2.3})$--$\mathcal{O}(N^{3})$ in the number of reactions \cite{orth2010flux,palsson2015systems}. Extensions incorporating regulatory logic or gene knockouts introduce combinatorial structure and render the problem NP-hard \cite{boyd2004convex,nocedal2006numerical}, necessitating heuristic or mixed-integer formulations. To capture temporal dynamics and regulatory feedback, kinetic models describe network evolution using nonlinear systems of ODEs, but their application at genome scale is limited by parameter uncertainty, stiffness arising from disparate timescales (systems where reaction rates span many orders of magnitude, forcing very small time steps), and the high per-timestep cost of implicit solvers. These challenges presage those encountered in whole-cell spatial modeling (\sectn\ref{subsec:model-wcm-spatial}), where spatial discretization further amplifies system dimensionality. When molecular copy numbers are low, stochastic formulations become essential: the chemical master equation (CME) provides a probabilistic description of network dynamics, with stochastic simulation algorithms (SSAs), such as Gillespie's algorithm \cite{gillespie1976general,gillespie1977exact}, sampling reaction trajectories at a computational cost proportional to the number of reaction channels ($R$) and total reaction events ($N_r$), $\mathcal{O}(N)$ where $N = RN_r$, and statistical convergence requiring $\mathcal{O}(1/\varepsilon^{2})$ samples with respect to the target error \cite{adalsteinsson2004biochemical,gillespie2007stochastic,mithani2009stochastic,clement2020stochastic}. Moreover, the time required to observe rare events or achieve mixing scales inversely with the spectral gap of the underlying Markov process (roughly how quickly the simulation reaches a representative statistical sample), $\mathcal{O}(1/\delta)$, posing a major bottleneck for networks with multiscale reaction kinetics. A detailed description of these classical network modeling approaches is provided in SI~\ref{sec:SI-wcm-network}. Together, these classical approaches form the computational backbone of network-level whole-cell models but face fundamental scalability barriers as system size, stochasticity, and coupling to spatial dynamics increase.

Quantum algorithms provide complementary acceleration pathways for metabolic and regulatory network modeling by targeting the linear algebra, optimization, and sampling kernels underlying classical formulations. Constraint-based metabolic models can be mapped to quantum convex optimization \cite{van2020convex, nannicini2024fast} and semidefinite programming solvers \cite{ brandao2017quantum}, yielding theoretical runtimes scaling as $\mathcal{O}(\sqrt{N}\,\mathrm{poly}(s,\kappa,1/\varepsilon))$ under oracle access assumptions. Here, $s$ is the sparsity of the matrix (the fraction of nonzero entries in the system matrix) and $\kappa$ its condition number (a measure of how sensitive the solution is to numerical error).
From a practical implementation perspective, modeling the reactions in a yeast cell using classical LP/QP solvers can require over 17 hours on an exascale supercomputer, whereas quantum solvers could, in principle, reduce this runtime to $\sim$10 secs under idealized assumptions (see SI~\ref{sec:SI-wcm-network} for a detailed estimate). 
Whereas, linear subproblems arising from steady-state balance constraints or Karush--Kuhn--Tucker (KKT) systems admit solution via QLSAs \cite{harrow2009quantum,morales2024quantum} with logarithmic dependence on problem dimension, $\mathcal{O}(\log N\, s^{2}\kappa \log(1/\varepsilon))$, for sparse and well-conditioned matrices, as well as quantum time evolution of the $S$ matrix \cite{kale2024simulation}. Combinatorial extensions of metabolic models can be formulated as quadratic unconstrained binary optimization (QUBO) problems and addressed using Grover-type search \cite{brisebois2025identifying}, variational approaches \cite{alevras2024mrna,tilly2022variational,roman2023quantum}, or the Quantum Approximate Optimization Algorithm (QAOA) \cite{blekos2024review}, offering quadratic speedups ($\mathcal{O}(\sqrt{N})$) over exhaustive classical search in favorable regimes. Stochastic network dynamics offer particularly strong opportunities for quantum advantage: amplitude-estimation-based or quantum Monte Carlo methods reduce sampling complexity from $\mathcal{O}(1/\varepsilon^{2})$ to $\mathcal{O}(1/\varepsilon)$ \cite{an2021quantum}, and quantum walk–based or quantum Markov chain algorithms accelerate mixing and rare-event statistics from $\mathcal{O}(1/\delta)$ to $\mathcal{O}(1/\sqrt{\delta})$ \cite{santha2008quantum,venegas2012quantum}. More direct quantum formulations encode the CME generator as a linear operator acting on the probability vector, enabling Hamiltonian simulation–based approaches with logarithmic dependence on the effective state-space dimension for suitably structured networks \cite{kazeev2014direct,kabengele2024modeling}. For solving a practical problem, such as the glycolysis pathway in \textit{E. coli}, classical pure SSA can take up to $\mathcal{O}(10^{13})$ operations per trajectory (42 days using an optimized implementation), whereas quantum approaches can potentially reduce this to just 15 operations per trajectory (see SI~\ref{sec:SI-wcm-network} for a detailed estimate). A comparative summary of classical and quantum computational scaling for network-level models is provided in \tbl\ref{tab:network_complexity}, with additional technical details discussed in SI~\ref{sec:SI-wcm-network}. Together, these methods suggest that hybrid quantum–classical workflows may substantially reduce the computational cost of large-scale network simulations by offloading the most expensive subroutines to quantum accelerators.

\begin{table}[t]
\centering
\caption{Computational complexity comparison of classical and quantum algorithms for metabolic and regulatory network modeling. Here $N$ is the number of variables (reactions, species, or fluxes), $s$ is matrix sparsity, $\kappa$ is the condition number, $\varepsilon$ is the target error, and $\delta$ is the spectral gap of the associated Markov chain. Quantum complexities assume oracle access and omit data-loading costs.}
\label{tab:network_complexity}
\begin{tabular}{p{4.2cm} p{3.5cm} p{4.4cm} p{2.4cm}}
\hline
\textbf{Problem} & \textbf{Classical complexity} & \textbf{Quantum complexity} & \textbf{Advantage} \\
\hline
Flux Balance Analysis (LP/QP) 
& $\mathcal{O}(N^{2.3})$--$\mathcal{O}(N^{3})$ 
& $\mathcal{O}\!\left(\sqrt{N}\,\mathrm{poly}(s,\kappa,1/\varepsilon)\right)$ (quantum convex / SDP solvers) 
& Polynomial in $N$ \\\\

Combinatorial structure
& NP-hard 
& $\mathcal{O}\!\left(\sqrt{N}\right)$ (QAOA, Grover's) 
& Exponential in $N$ \\\\

Stochastic simulation (SSA sampling) 
& $\mathcal{O}(N)$ 
& $\mathcal{O}\!\left(\log Ns^{2}\kappa\log(1/\varepsilon)\right)$ (QLSA) 
& Exponential in $N$ \\\\

Stochastic simulation (target error) 
& $\mathcal{O}(1/\varepsilon^{2})$ 
& $\mathcal{O}(1/\varepsilon)$ (quantum Monte Carlo / amplitude estimation) 
& Quadratic in $1/\varepsilon$ \\\\

Markov chain mixing and Rare-event 
& $\mathcal{O}(1/\delta)$ 
& $\mathcal{O}(1/\sqrt{\delta})$ (quantum walks) 
& Quadratic in $1/\sqrt{\delta}$ \\

\hline
\end{tabular}
\end{table}

The progression from network models to spatial whole-cell simulations (\sectn\ref{subsec:model-wcm-spatial}) represents a natural extension: the kinetic ODE and stochastic frameworks described here are embedded within spatially resolved reaction-diffusion equations, multiplying the system dimension by the number of spatial voxels. Quantum advantages demonstrated at the network level, particularly for stochastic sampling and large linear system solves, thus provide foundational building blocks for quantum-enhanced whole-cell spatial modeling.

\begin{tcolorbox}[colframe=black, colback=ORNLgreen!60, sharp corners=southwest, boxrule=1pt, width=\textwidth, left=2mm, right=2mm, top=1mm, bottom=1mm]
\label{box:practical_network}
\textbf{State of the art --- regulatory-network logic on quantum hardware.}
A five-protein regulatory network governing mammalian cortical neurodevelopment was resolved on IBM gate-based hardware by casting its Boolean decisional logic as a B-SAT problem and solving the classical NP-hard problem with Grover's algorithm at $\mathcal{O}(\sqrt{N})$ computational complexity~\cite{brisebois2025identifying}. This demonstrates quantum search on real hardware for the NP-hard combinatorial kernel that arises when regulatory structure is added to network models. We highlight state-of-the-art ODE solutions on quantum hardware in \sectn\ref{subsec:model-wcm-spatial}.
\end{tcolorbox}

\subsection{Whole-cell spatial modeling}\label{subsec:model-wcm-spatial}

Spatial modeling of the whole cell requires capturing coupled biochemical, mechanical, and transport processes operating across wide ranges of spatial and temporal scales. Johnson et al.~\cite{johnson2021quantifying} provide a comprehensive overview of numerical approaches used to study these systems. At one extreme, molecular-scale descriptions rely on atomistic simulations such as molecular dynamics, resolving motions on femtosecond to microsecond timescales ($10^{-15}$--$10^{-7}$~s), but these methods are computationally intractable for whole-cell applications \cite{stevens2023molecular}. At the cellular scale, biologically relevant dynamics such as growth, division, and intracellular transport unfold over timescales spanning microseconds to hours ($10^{-6}$--$10^{3}$~s), representing an $\mathcal{O}(10^{9})$ separation in temporal scales. Practical whole-cell models therefore adopt either non-spatial formulations, which track the temporal evolution of species concentrations (of metabolites and proteins), or spatially resolved formulations, which additionally account for the movement and spatial organization of the species within the cell.

Classical numerical techniques for whole-cell spatial modeling encompass both deterministic and stochastic approaches in non-spatial and spatial settings. Deterministic models assume sufficiently high molecular copy numbers and smooth concentration fields, leading to systems of coupled ordinary and partial differential equations (ODEs and PDEs) describing reaction--diffusion processes, electrostatics (Poisson equations), and cellular mechanics. Solving these linear and nonlinear ODEs and PDEs using implicit methods typically incurs computational costs scaling as $\mathcal{O}(N^{1.5})$--$\mathcal{O}(N^{3})$ with respect to the system size~\cite{coppersmith1982asymptotic,trefethen2022numerical}. In contrast, stochastic formulations explicitly model thermal fluctuations, molecular collisions, and discrete reaction events, providing a more faithful description when copy numbers are low \cite{ghaemi2020silico,das2021high}. These approaches involve sampling trajectories of the CME or solving stochastic PDEs, with computational complexity scaling as $\mathcal{O}(1/\varepsilon^{2})$ for statistical convergence and $\mathcal{O}(1/\delta)$ for Markov chain mixing and rare-event sampling. Practically, hybrid stochastic-deterministic simulations offer a combination of scalability and capability to capture the realistic phenomenology of the species \cite{thornburg2022fundamental,thornburg2026bringing}.

Quantum approaches to whole-cell spatial modeling target the dominant linear algebra, sampling, and mixing bottlenecks in both deterministic and stochastic formulations. For linear ODEs and PDEs arising from reaction--diffusion systems, QLSAs~\cite{harrow2009quantum,morales2024quantum} and Schr\"odingerization-based quantum simulations~\cite{hu2024quantum} offer logarithmic scaling with respect to the system size, $\mathcal{O}(\log N\, s^{2}\kappa \log(1/\varepsilon))$, for sparse and well-conditioned operators. In reality, the approach for encoding the states of the PDEs into the quantum computer can be a bottleneck for the computational scaling of these algorithms. We elaborate more on the challenges of data encoding in \sectn\ref{sec:discussion}. From a practical implementation perspective, modeling a single cell cycle of a yeast cell using classical linear PDE solvers can require over 26 days on an exascale supercomputer, whereas quantum linear solvers could, in principle, reduce this runtime to $\sim$5 hours under idealized assumptions (see SI~\ref{sec:SI-wcm-spatial} for a detailed estimate). While quantum systems are inherently linear, nonlinear PDEs may be addressed by transforming the nonlinear PDEs to be linear using techniques \cite{leyton2008quantum,lloyd2020quantum} such as Carleman linearization~\cite{liu2021efficient}, Koopman--von Neumann formulations~\cite{joseph2020koopman}, or level-set methods~\cite{jin2024quantum}. A recent effort has used Careleman linearization to decompose reaction-diffusion equations in subcellular systems into a linear combination of Hamiltonian simulations \cite{lockwood2025quantum}. Moreover, methods that directly encode nonlinear dynamics can be used, such as quantum amplitude estimation algorithm (QAEA) to solve specific nonlinear PDEs \cite{gaitan2020finding}, Kolmogorov equation approximation to the limit of large linear combination of operations \cite{bravyi2025quantum}, and tensor network–based algorithms \cite{lubasch2018multigrid,lubasch2020variational,meena2024towards,gopalakrishnan2026tensor} with computational complexity scaling as $\mathcal{O}(\log N d^{2})$--$\mathcal{O}(\log N d^{3})$, where $d$ is the bond dimension of the tensor network. Stochastic spatial models can be accelerated using quantum Monte Carlo and amplitude estimation techniques~\cite{an2021quantum,jin2025quantum}, reducing sampling complexity to $\mathcal{O}(1/\varepsilon)$, and quantum Markov chain or quantum walk–based algorithms~\cite{szegedy2004quantum,layden2023quantum,santha2008quantum,venegas2012quantum}, which improve mixing and rare-event scaling to $\mathcal{O}(1/\sqrt{\delta})$. For the stochastic spatial counterpart, modeling a single \textit{E.~coli} cell cycle using a classical SSA-based RDME solver requires $\sim$92 years on a single-core workstation and $\sim$13 hours to 5.5 days on an exascale HPC system even with ideal domain decomposition, whereas quantum Markov chain and quantum walk-based algorithms could, in principle, reduce this to $\sim$0.02~seconds (see SI~\ref{sec:SI-wcm-spatial} for a detailed estimate). A comparative summary of classical and quantum computational complexity for whole-cell spatial models is provided in Table~\ref{tab:wholecell_complexity}.

\begin{table}[t]
\centering
\caption{Computational complexity comparison of classical and quantum algorithms for deterministic and stochastic whole-cell spatial models. $N$ denotes the number of coupled degrees of freedom after discretization (species, reactions, total number of spatial voxels), $s$ is sparsity, $\kappa$ is the condition number, $\varepsilon$ is the target accuracy, $\delta$ is the spectral gap of the induced Markov process, and $d$ is the bond dimension for the tensor network approach.}
\label{tab:wholecell_complexity}
\begin{tabular}{p{4.2cm} p{3.5cm} p{4.4cm} p{2.4cm}}
\hline
\textbf{Model type} & \textbf{Classical complexity} & \textbf{Quantum complexity} & \textbf{Advantage} \\
\hline
Deterministic non-spatial ODEs and spatial PDEs (reaction--diffusion) 
& $\mathcal{O}(N^{1.5})-\mathcal{O}(N^{3})$ per timestep (implicit solvers) 
& $\mathcal{O}\!\left(\log Ns^{2}\kappa\log(1/\varepsilon)\right)$ (QLSA) 
& Exponential in $N$ \\\\

Nonlinear PDEs 
& $\mathcal{O}(N^{1.5})-\mathcal{O}(N^{3})$
& $\mathcal{O}\!\left(\log Ns^{2}\kappa\log(1/\varepsilon)\right)$ (linearization-based) and $\mathcal{O}\!\left(\log Nd^2\right)-\mathcal{O}\!\left(\log Nd^3\right)$ (tensor-network methods)
& Exponential in $N$ (method dependent) \\\\

Stochastic simulation (SSA sampling) 
& $\mathcal{O}(N)$ 
& $\mathcal{O}\!\left(\log Ns^{2}\kappa\log(1/\varepsilon)\right)$ (QLSA) 
& Exponential in $N$ \\\\

Stochastic simulation (target error) 
& $\mathcal{O}(1/\varepsilon^{2})$ 
& $\mathcal{O}(1/\varepsilon)$ (quantum Monte Carlo) 
& Quadratic in $1/\varepsilon$ \\\\

Stochastic simulation (mixing / rare events) 
& $\mathcal{O}(1/\delta)$ 
& $\mathcal{O}(1/\sqrt{\delta})$ (quantum Markov chains) 
& Quadratic in $1/\sqrt{\delta}$ \\

\hline
\end{tabular}
\end{table}

\begin{tcolorbox}[colframe=black, colback=ORNLgreen!60, sharp corners=southwest, boxrule=1pt, width=\textwidth, left=2mm, right=2mm, top=1mm, bottom=1mm]
\label{box:practical_spatial}
\textbf{State of the art --- differential equations on quantum hardware.}
A 5{,}043-dimensional linear system from an unsteady acoustic-wave PDE on a superconducting processor, using an error-suppressed iterative QLSA scaled up by a classical subspace method~\cite{chen2024enabling}. Related hardware demonstrations span small linear systems in fluid dynamics and finance~\cite{yalovetzky2024solving,gopalakrishnan2024solving,song2025incompressible,lu2025practical,wang2026simulating}. While no hardware demonstration yet exists for whole-cell spatial modeling, these results show that the linear-system and PDE kernels underlying spatial whole-cell modeling can already be executed end-to-end on real quantum hardware.
\end{tcolorbox}


\section{Discussion and Conclusion}\label{sec:discussion}

Quantum computational approaches hold significant promise for accelerating simulations across the three hierarchical scales that underpin cellular modeling: atomistic–molecular interactions, biochemical networks, and whole-cell spatial dynamics. In this work, we have examined how quantum algorithms may advance each of these domains, supported by a systematic complexity analysis that highlights regimes where substantial theoretical speedups over classical methods may be achievable. At present, however, these advantages remain largely prospective, and progress is expected to continue within individual disciplinary boundaries in the near term. The most transformative advances will likely emerge from the integration of these scales, linking biomolecular simulations with network-level dynamics embedded in spatially resolved whole-cell models. Achieving such multiscale integration would mark a critical step toward understanding how molecular-level events give rise to cellular function, ultimately enabling deeper insight into therapeutic responses, environmental influences, and the behavior of complex biological systems.

While there are multiple avenues through which quantum computing could theoretically enhance whole-cell modeling, several critical considerations remain for the practical realization of these approaches. Foremost among these is the challenge of integrating the disparate systems explored here into a unified, multiscale framework. Progress toward this goal will require overcoming key hurdles, particularly in the incorporation of experimental data into computational models. Growing information in structural databases have enabled increasingly detailed biomolecular simulations, while ``omics" datasets, including genomics, transcriptomics, and proteomics, have driven the development of biochemical network models. In parallel, diverse imaging modalities are informing dynamic, spatially resolved whole-cell representations. Despite these advances, available data remain incomplete, especially when accounting for the full diversity of biomolecules and cellular processes such as transport, regulation, and cell division. In this context, quantum computing may play a complementary role by accelerating the estimation of missing parameters, similar to emerging AI/ML approaches; for instance, quantum-enhanced MD could improve predictions of molecular binding and reaction kinetics that underpin network behavior. Moving toward quantum-enabled whole-cell simulations will also require addressing substantial technical challenges, including efficient input and output of data to quantum systems, the development of hybrid classical–quantum computing platforms, and continued advances in algorithms and hardware. These computational challenges are considered in greater detail below.

\subsection{Pathway for integration of models: A hybrid quantum-HPC approach}\label{subsec:discussion-integration}

\begin{figure}[ht!]
\centering
\includegraphics[width=0.95\textwidth]{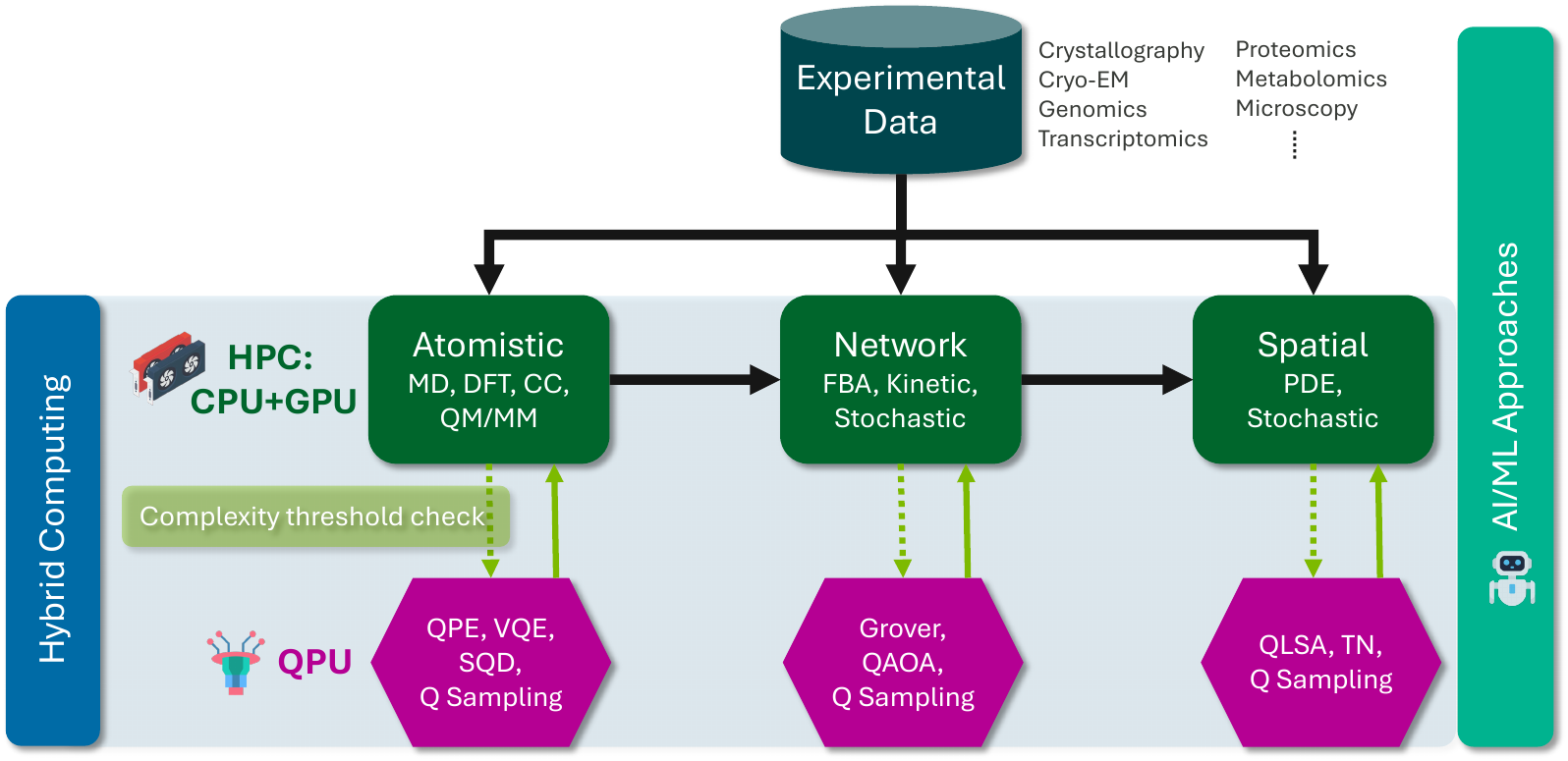}
\caption{Depiction of a hybrid HPC-Quantum-AI integration framework to enhance various hierarchical components of whole-cell modeling. AI/ML approaches can span across the whole modeling paradigm, from experimental data to surrogate modeling and AI for quantum algorithms and computing. The $\cdots$ arrows to the QPUs denote the check for complexity threshold to offload certain kernels from the HPC solvers to the QPU.}\label{fig:integration}
\end{figure}

An integrated quantum-only implementation of all the three modeling aspects can be challenging due to: (1) the limited data input for atomistic-molecular systems; (2) missing models of micro- and macro-scale networks \cite{orth2010systematizing,hosseini2017discovering}; and (3) data encoding limitations for extreme-scale spatial models. Thus, a more realistic integration approach would revolve around a hybrid quantum-HPC (Q-HPC) framework \cite{beck2024integrating,shehata2026bridging}.  A visual depiction of this unified framework is shown in \fig\ref{fig:integration}. The main objective would be to develop a unified HPC solver with a quantum processing unit (QPU) as an accelerator to offload certain kernels within the HPC solver, similar to how graphic processing units (GPUs) are currently used in whole-cell modeling \cite{thornburg2026bringing}. Furthermore, current state-of-the-art AI/ML can be incorporated as surrogate models to compliment the computational acceleration intended by GPU and QPU offloading. Moreover, AI/ML approaches can supplement various aspects of the whole modeling framework, from experimental data curation to modeling efforts for quantum algorithms and computing. Unlike the general use-case of GPUs, QPUs can be used only with certain constraints, e.g., special oracles are needed for different kernels and observations (outputs) extracted from circuits are limited to expectation values or samples, rather than full state vectors. Here, we provide some potential avenues where QPUs can be used in the three modeling approaches.

For the atomistic models, the main MD code can remain classical, whereas at each step of the MD simulation, QC can be used to: (1) compute energies and forces on an active region, and (2) perform quantum-accelerated sampling. For network modeling, quantum optimization and sampling-based algorithms can be used to solve both sub- or whole-network models. Finally, for the whole-cell models, QLSA can be used to solve large, stiff linear systems of equations at each time step or in an implicit solver. Additionally, tensor network-based algorithms can be used when quantum entanglement is low (i.e., the bond dimension $d$ remains tractable), and quantum Monte Carlo and walk-based algorithms for stochastic sampling simulations. While most of these algorithms do promise theoretical computational advantage compared to classical counterparts, for practical implementations, we should rely on them only when needed --- when a certain threshold of problem complexity or size is exceeded during the simulation. On other occasions, the classical techniques would prove to be more efficient considering the data input-output (I/O) and latency challenges of using a QPU.

One potential general procedure for the integrated framework could be as follows. (1) Identify the kernels that can be called (or offloaded) on the QC, such as quantum chemistry, stochastic sampling, optimization, and linear/nonlinear differential equation solvers. (2) For a given kernel, add a threshold for a complexity level, e.g., FBA on a $10^{10}$ variable LP for a given condition (see Yeast system in \fig~\ref{fig:time_complexity_network-FBA} and \tbl~\ref{tab:wcm-network_FBA_times}), a large-scale implicit PDE solve, and reaction kinetics within a single voxel. This is when the kernel can be offloaded to the QPU, expecting a computational advantage. (3) The quantum solution, which can be samples or a small number of observables requiring fewer measurements than full state tomography, is extracted and looped back into the main HPC workflow.

While conceptually the hybrid Q-HPC framework could be a more practical approach to integrate the quantum algorithms into the whole-cell modeling paradigm, we still need to address the practical bottlenecks faced by the quantum computing community. This includes not only hardware challenges, but also algorithmic and software developments required.

\subsection{Practical challenges for implementation}\label{subsec:discussion-practical}

\subsubsection{Algorithmic challenges and opportunities} 
The potential computational advantages for the quantum algorithms described in \sectn\ref{sec:modeling} are dependent on various assumptions, including the availability of an efficient quantum oracle, the omission of data-loading costs, and the use of expectation values as observables (measuring only a single qubit) instead of full-state tomography (measuring the states of all or most qubits). In reality, for problems that are not intrinsically quantum mechanical, such as solving linear systems of equations for a continuous system discretized by grids/voxels, the potential speed-ups require solving extremely large system sizes (see discussion of whole-cell simulations in SI~\ref{sec:SI-wcm-spatial}). Encoding such datasets onto quantum devices can be extremely challenging, potentially limiting the attainable quantum advantage of the full framework \cite{hoefler2023disentangling}. An alternative is to tackle localized problems (within each voxel or a subset of voxels) or distribute the problem over many QPUs \cite{kim2025distributedvariationalquantumalgorithm}. Furthermore, instead of measuring the full state, targeting kernels that only require expectation values of the state for QPU offloading would be needed for efficient tomography, such as models requiring only the overall status of the cell.

Beyond adapting existing quantum algorithms to biological problems, several algorithmic challenges specific to whole-cell modeling may motivate the development of new quantum methods. First, current metabolic and regulatory network models are necessarily incomplete: many interaction partners, regulatory edges, and kinetic parameters remain uncharacterized. Quantum-enhanced machine learning and optimization approaches could assist in inferring missing network components from high-dimensional omics datasets, complementing classical systems-identification methods. Second, the governing PDEs for spatial whole-cell dynamics are typically derived from phenomenological rate laws rather than first-principles Hamiltonians. Quantum algorithms that can learn or infer effective Hamiltonians directly from trajectory data, leveraging quantum versions of Koopman operator methods~\cite{joseph2020koopman} or Hamiltonian learning protocols~\cite{lockwood2025quantum}, could provide more physically grounded dynamical models. Both directions represent opportunities where the quantum computing and systems biology communities could develop co-designed algorithms that do not simply port classical formulations onto quantum hardware, but instead exploit quantum representations natively.

\subsubsection{Incorporating experimental data into the algorithms} 
Another key data I/O challenge concerns the availability of experimental data that can be effectively utilized in the various quantum implementations. For the atomistic-molecular models, for which accurate ground states are required for quantum simulations, data are typically obtained from electron, neutron, or X-ray scattering experiments to characterize the various states of matter being modeled. For the whole-cell network models, baseline network interactions from omics data and reaction networks identified by experimental trials are critical. For the spatial whole-cell models, the initial and boundary conditions of the various cellular components are obtained through imaging techniques. Crucially, the outputs from quantum implementations at each scale should feed forward to inform the experimental inputs and initial conditions at the next scale: for example, quantum-computed ground-state energies can refine reaction rate parameters in network models, which in turn constrain the initial conditions for spatial simulations. A clear synergy between computational models and experimental observations is therefore required to effectively deploy quantum counterparts of the classical whole-cell modeling paradigm.

\subsubsection{Hardware requirements and timeline for practical implementation}
Another assumption for many of the quantum algorithms is that they require fault-tolerant quantum computers (FTQC) in order to simulate large circuits without hampering the prediction accuracies (fidelity). Achieving fault-tolerant quantum computation requires continued progress in quantum error correction. As FTQC hardware matures, the individual quantum kernels identified here will need to be implemented, validated, and benchmarked on real devices. In the current era of noisy intermediate-scale quantum (NISQ) computing, variational algorithms \cite{cerezo2021variational} pose as a viable option for practical implementation and demonstration of the hybrid quantum-HPC frameworks. Having a hybrid software framework amenable to both standalone quantum and variational quantum algorithms is thus crucial \cite{shehata2026bridging}.

These challenges on both algorithmic implementation and hardware capability raise the question of when such quantum enhancements for whole-cell modeling can be practically achieved. For estimating this timeline, from an algorithmic perspective, one needs to carefully choose the kernels for which a quantum offload would result in the best computational advantage (speed-up or data reduction/compression). Generally, quantum algorithms which offer exponential speedup compared to their classical counterparts are foreseen to demonstrate immediate quantum utility \cite{hoefler2023disentangling}. Thus, some of the candidate kernels could be quantum simulation to obtain the ground state energy, solving the combinatorial structure of a gene regulatory network, or solving stiff PDEs in the spatial whole-cell model. Nonetheless, convergence and sampling in stochastic simulations, for both the network models and high-dimensional spatial models, are a key computational bottleneck for the scales of problems in the whole-cell context and could also be potential candidates for offloading.

The next key component for estimating the timeline of quantum utility is the hardware --- how many fault-tolerant qubits would one need and how many operations (gate depth) can one accommodate? Hardware requirements vary substantially across the three modeling scales considered here. For atomistic-molecular simulations, Reiher et al.~\cite{reiher2017elucidating} estimated that elucidating the reaction mechanism of nitrogen fixation in nitrogenase, a prototypical enzyme-scale quantum chemistry problem, would require on the order of $10^{14}$ $T$-gate operations on a fault-tolerant device with $\sim$2,000 logical qubits, far beyond current hardware but within the scope of projected longer-term roadmaps. More recently, Alexeev et al.~\cite{alexeev2025perspective} surveyed quantum chemistry applications tractable with 25--100 logical qubits, identifying molecular energy estimation, reaction pathway analysis, and small-molecule binding as near-term targets. Scaling from molecular to network and spatial whole-cell problems will require substantially more resources: QLSA-based solvers for the spatial PDE systems considered in SI~\ref{sec:SI-wcm-spatial} scale as $\mathcal{O}(\log N \kappa s^2)$ in circuit depth, but with $N \sim 10^{11}$--$10^{16}$ for whole-cell spatial models, even logarithmic scaling translates to circuits of non-trivial depth, and the large condition numbers ($\kappa$) expected for stiff biological PDEs will further  amplify qubit and gate requirements. Establishing concrete resource estimates for the full range of kernels identified here, from FBA optimization to RDME sampling, remains an open and important direction.

\subsection{Final remarks}\label{subsec:discussion-final}

In pursuit of the goal of going from atoms to the basic biological unit of life, a cell, this \perspective has identified a hierarchy of quantum algorithmic opportunities, from near-term variational approaches for molecular-scale quantum chemistry to fault-tolerant QLSA and quantum walk-based methods for spatial stochastic whole-cell models, and grounded these in concrete computational estimates for representative biological systems. The path forward will require co-development across the quantum algorithms, quantum hardware, and biological modeling communities: quantum kernels must be benchmarked on real devices against classical baselines, fault-tolerant qubit counts must be projected for biologically relevant problem sizes, and hybrid Q-HPC software frameworks must be designed to accommodate the data I/O constraints imposed by quantum measurement. We anticipate that a staged integration strategy, beginning with quantum-accelerated molecular simulations and progressing toward network-scale and ultimately spatial whole-cell models as hardware matures, offers the most tractable near-term pathway. In this framework, QPUs would serve as targeted accelerators for selected molecular, optimization, sampling, and linear-algebra kernels. Classical HPC resources would remain responsible for much of the broader multiscale workflow.

Once thoroughly validated against experiments and classical benchmarks, quantum-enabled whole-cell simulation could connect atomic-scale events, such as molecular recognition, conformational changes, and reaction energetics, to cellular observables including metabolic state, growth, stress responses, and phenotypic variability. This capability could help identify emergent behavior arising from the collective interactions of cellular components. It could also enable systematic in silico perturbation studies that may be difficult, expensive, or impossible to perform experimentally. Quantum-enabled simulations would complement structural, omics, and imaging experiments by helping generate and prioritize mechanistic hypotheses. They could predict the effects of mutations, environmental changes, or molecular interventions. They could also identify experiments that most effectively distinguish among competing models.

Validated quantum-enabled simulations could also support AI/ML approaches. High-fidelity molecular and multiscale simulations could generate physically constrained training data for surrogate models that approximate expensive simulation components, such as reaction energetics, rate parameters, or local spatial dynamics. These AI models could then accelerate repeated whole-cell simulations, parameter inference, and uncertainty analysis within hybrid Q-HPC workflows. Such training data would not replace experimental datasets. Rather, they could augment them where direct measurements are sparse, difficult to obtain, or limited in resolution. 

Realizing this vision would mark a transformational advance: for the first time, the full molecular complexity of a living cell, resolved in space and time (\fig\ref{fig:outline}), could become computationally accessible. Such a development offers the prospect of better explaining how cellular complexity emerges as one traverses from the smallest spatial and temporal scales to those that are relevant for life itself.


\section*{Data and code availability}

The specific codes and data that support the findings of this study will be made openly available after peer review.

\section*{Acknowledgments}
This research used resources of the Oak Ridge Leadership Computing Facility at the Oak Ridge National Laboratory, which is supported by the Advanced Scientific Computing Research programs in the Office of Science of the U.S. Department of Energy under Contract No. DE-AC05-00OR22725.


\section*{Author declaration}
The authors declare no competing interests.

\bibliographystyle{ieeetr}
\bibliography{refs.bib}

\clearpage
\appendix


\section*{Supplementary Information}
\label{S1_Appendix}

\renewcommand{\thepage}{S-\arabic{page}}
\setcounter{page}{1}
\renewcommand{\thefigure}{S-\arabic{figure}}
\setcounter{figure}{0}
\renewcommand{\thetable}{S-\arabic{table}}
\setcounter{table}{0}
\renewcommand{\theequation}{S-\arabic{equation}}
\setcounter{equation}{0}

\section{Computing hardware consideration for scaling analysis}\label{sec:SI-scaling-hardware}

For the classical approaches, we consider a personal computer or workstation (PC) (with 500 GigaFLOPS of compute performance) and an HPC system (OLCF's Frontier supercomputer with 1.6 ExaFLOPS \cite{atchley2023frontier}) as the hardware used. For the quantum hardware, we assume an ideal hardware that will offer a throughput of 10.5 kop/s (kilo-operations per sec) with 16-bit floating point precision \cite{hoefler2023disentangling}. We note that the common metric for reporting performance (calculation speed) of a quantum computer is Circuit Layer Operations per Second (CLOPS) \cite{wack2021qualityspeedscalekey}. The IBM Quantum Heron processor is reported to have over 330 kCLOPS and IQM's 20-qubit hardware is reported to have 2.6 kCLOPS \cite{abdurakhimov2024technology}. Because CLOPS cannot be directly compared to FLOPS (Floating Point Operations per Second) as each circuit layer can have multiple gate operations, we use the metric proposed in \cite{hoefler2023disentangling} for comparison of the algorithm performance.

\section{Whole-cell metabolic and regulatory network modeling: sample scaling analysis}\label{sec:SI-wcm-network}

\subsection{Constraint-based models}\label{subsec:SI-wcm-network_FBA}

Genome-scale metabolic models (GEMs) represent the most widely adopted framework for network-level modeling. Construction begins with genome annotation, mapping genes to enzymatic functions using curated databases such as KEGG, BioCyc, and ModelSEED. Draft models are assembled using template reconstructions from well-characterized reference organisms, followed by gap-filling algorithms that identify minimal reaction sets needed to restore network connectivity. The resulting stoichiometric matrix $S$, encoding mass balance relationships between metabolites (rows) and reactions (columns), forms the foundation for constraint-based analysis.

Flux balance analysis (FBA) formulates metabolism as a linear optimization problem: given steady-state constraints ($S\mathbf{v} = 0$), flux bounds, and an objective function (typically biomass maximization), the optimal flux distribution is computed via linear programming (LP). Extensions include parsimonious FBA (pFBA), which minimizes total flux through quadratic programming (QP), and regulatory FBA (rFBA), which integrates gene expression data to impose condition-specific constraints. The computational complexity of these methods scales as $\mathcal{O}(N^{2.3})$ to $\mathcal{O}(N^3)$ for LP/QP solvers using the simplex or interior-point methods \cite{orth2010flux,palsson2015systems}, where $N$ is the number of reactions. For combinatorial extensions such as OptKnock (identifying gene knockouts for metabolic engineering), the problem becomes NP-hard \cite{boyd2004convex,nocedal2006numerical}, requiring heuristic or mixed-integer approaches.

\begin{figure}[!ht]
\centering
G\includegraphics[width=0.9\textwidth]{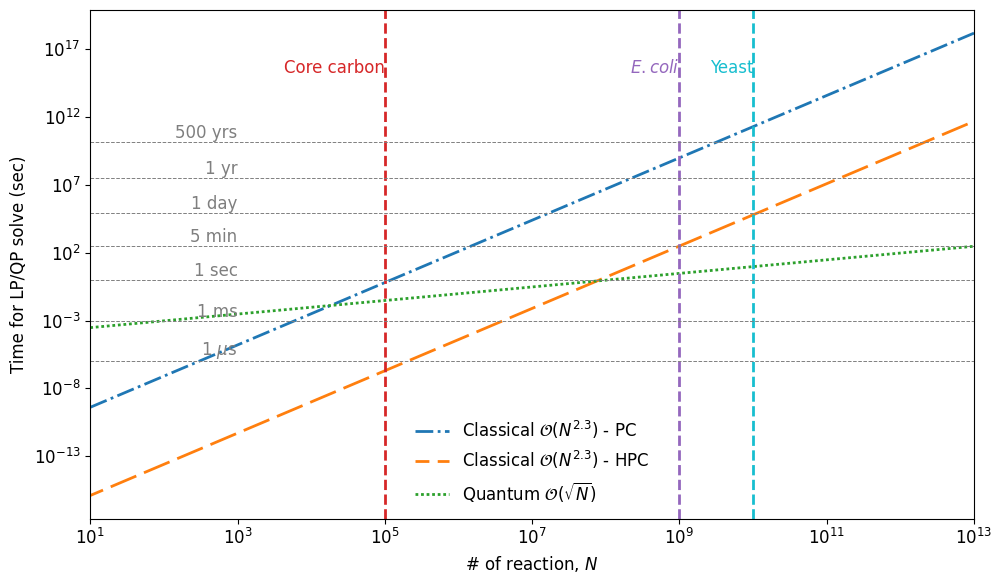}
\caption{\textbf{Classical and quantum FBA solver time complexity study.} Time complexity of different linear optimization solvers for solving the constrained flux balance equations with increasing system size (number of reactions solved). For the classical solution, time taken by the solver on an HPC system (1.6~ExaFLOPS) and a personal computer (500~GigaFLOPS) are shown. The quantum system is assumed to have a throughput of 10.5~kop/s of 16-bit floating point precision \cite{hoefler2023disentangling}. Reference vertical lines represent the system sizes for core carbon metabolism and genome-scale \textit{E. coli} (iML1515) and Yeast.}
\label{fig:time_complexity_network-FBA}
\end{figure}

Figure \ref{fig:time_complexity_network-FBA} depicts the time scalability of classical and quantum linear optimization solvers to tackle various practical FBA problems. We consider three practical problems of varying complexity: core carbon metabolism with glycolysis, TCA, and PPP \cite{orth2010reconstruction}; genome-scale \textit{E. coli} (iML1515) \cite{monk2017ml1515}; and genome-scale Yeast \cite{lu2019consensus}. For these three systems, we consider $[95, 2712, 4058]$ reactions, $[75, 1877, 2742]$ metabolites, and $[80, 1516, 1150]$ genes, giving a total of $\mathcal{O}(10^5)$, $\mathcal{O}(10^9)$, and $\mathcal{O}(10^{10})$ reactions, respectively. Figure~\ref{fig:time_complexity_network-FBA} depicts the scaling of the computational complexity of classical LP/QP solvers (using a PC and an exascale HPC system) and a quantum solver. A summary of the computational cost for solving the three problems by the classical and quantum solvers is provided in \tbl\ref{tab:wcm-network_FBA_times}.  While the quantum approach may not offer significant computational speedup over the HPC solution for the core carbon metabolism, it helps accelerate the \textit{E. coli} and Yeast computations.

\begin{table}[!ht]
\centering
\caption{Estimated wall-clock time to solve the FBA LP for three model systems of increasing size using classical (PC and HPC) and quantum solvers. Classical complexity scales as $\mathcal{O}(N^{2.3})$; quantum complexity scales as $\mathcal{O}(\sqrt{N})$ \cite{nannicini2024fast,van2020convex}. The PC is assumed to operate at 500~GigaFLOPS and the HPC at 1.6~ExaFLOPS (OLCF Frontier \cite{atchley2023frontier}); the quantum system is assumed to deliver 10.5~kop/s of 16-bit floating-point throughput \cite{hoefler2023disentangling}. $N$ denotes the total number of reactions (variables) in the LP.}
\label{tab:wcm-network_FBA_times}
\setlength{\tabcolsep}{8pt}
\begin{tabular}{lrrrr}
\hline
\textbf{System} & $N$ & \textbf{PC} & \textbf{HPC (Frontier)} & \textbf{Quantum} \\
\hline
Core carbon (glycolysis, TCA, PPP)
  & $10^{5}$  & $0.63$~s        & $0.20~\mu$s             & $30.1$~ms \\
\textit{E.~coli} (iML1515)
  & $10^{9}$  & $31.8$ years    & $5.2$~min               & $3.01$~s  \\
Yeast (Yeast8)
  & $10^{10}$ & $6{,}341$ years & $17.4$~hrs              & $9.52$~s  \\
\hline
\end{tabular}
\end{table}

\subsection{Kinetic and dynamic models} \label{subsec:SI-wcm-network_kinetic}

While FBA captures steady-state behavior, cellular responses to perturbations, transient dynamics, and regulatory feedback require kinetic models formulated as systems of ODEs:
\begin{equation}
    \frac{d\mathbf{c}}{dt} = S \cdot \mathbf{v}(\mathbf{c}, \mathbf{p})
\end{equation}
where $\mathbf{c}$ is the vector of metabolite concentrations, $\mathbf{v}(\mathbf{c}, \mathbf{p})$ contains rate laws (e.g., Michaelis-Menten, Hill kinetics) parameterized by $\mathbf{p}$, and $S$ is the stoichiometric matrix. Unlike FBA, kinetic models require experimentally measured or estimated parameters ($K_m$, $V_{max}$, binding constants) which are often unavailable or poorly constrained, limiting model scope.

For small to medium networks (tens to hundreds of reactions), these ODE systems can be solved using standard numerical integrators. However, genome-scale kinetic models face severe computational challenges: stiff equations arising from disparate timescales (fast binding equilibria versus slow gene expression), nonlinear rate laws requiring implicit solvers with $\mathcal{O}(N^{1.5})$ to $\mathcal{O}(N^3)$ complexity per timestep, and parameter uncertainty necessitating ensemble simulations or sensitivity analyses that multiply computational burden. These challenges presage those encountered in whole-cell modeling (\sectn\ref{subsec:model-wcm-spatial}), where spatial discretization further amplifies system dimensionality.

\subsection{Stochastic network models}\label{subsec:SI-wcm-network_stochastic}

When molecular copy numbers are low---as in gene regulatory circuits where transcription factors may number only tens of molecules per cell---deterministic ODE models break down and stochastic fluctuations become functionally significant. The chemical master equation (CME) provides a rigorous probabilistic framework, describing the time evolution of the probability distribution over discrete molecular states. Stochastic simulation algorithms (SSA), such as Gillespie's algorithm \cite{gillespie1976general,gillespie1977exact}, sample trajectories from this distribution by iteratively selecting reaction events according to their propensities \cite{adalsteinsson2004biochemical,gillespie2007stochastic,mithani2009stochastic,clement2020stochastic}.

The computational cost of SSA scales with the number of reaction channels ($R$) and total reaction events ($N_r$), yielding $\mathcal{O}(N)$ complexity where $N = R \times N_r$. Convergence to accurate statistics requires $\mathcal{O}(1/\varepsilon^2)$ samples for target error $\varepsilon$, while rare event sampling depends on the spectral gap $\delta$ of the underlying Markov chain, scaling as $\mathcal{O}(1/\delta)$. For networks with fast and slow reactions spanning orders of magnitude in timescale, hybrid tau-leaping or multi-scale approaches offer computational savings at the cost of approximation accuracy.

To estimate the computational complexity for practical problems, we consider a gene regulatory circuit and glycolysis pathway as two candidate systems. For the gene regulatory circuit, the total propensity is taken as $a_0=230$ events/min. For \textit{E. coli} with a cell cycle of $T_\text{sim} = 20$ min, this results in $N_r= a_0 \times T_\text{sim} = 4,600$ events per trajectory (or events per replicate). Considering $R=20$ reaction channels, the total operations per trajectory for classical SSA solver is $N=R\times N_r=\mathcal{O}(10^5)$, whereas $\mathcal{O}(\log N)=5$ operations per trajectory using a quantum linear solver (assuming ideal system sparsity and condition number). For the glycolysis pathway, we consider a total propensity of $a_0=10^9$ events/sec, giving $N_r=1.2\times 10^{12}$ events for a cell cycle of \textit{E. coli}, and with $R=30$ reaction channels, we end up with $N = \mathcal{O}(10^{14})$ operations per trajectory. Using an optimized single-core C++ implementation (e.g., StochKit2 \cite{sanft2011stochkit2}, achieving approximately $10^7$ reaction events/sec on a modern workstation CPU), a single trajectory of the glycolysis pathway could take up to 42 days. On a leadership-class HPC system, ensemble parallelism across concurrent trajectories can reduce wall-clock time proportionally for statistical convergence; however, the per-trajectory cost remains unchanged since SSA is inherently serial. Using a quantum linear solver, the computational complexity can be reduce to around 14 operations per trajectory. 

For statistical convergence of these simulations, say to observe the mean protein count for the gene regulatory circuit, with a relative error of $\varepsilon=5\%$, one would need about $\mathcal{O}(1/\varepsilon^2) = 400$ samples (or replicates) with classical Monte Carlo. This can increase up to $10,000$ samples if one requires a target error of $1\%$. Whereas, using quantum Monte Carlo ($\mathcal{O}(1/\varepsilon)$), the number of samples can reduce to around 20 and 100 for $5\%$ and $1\%$ target error, respectively. Finally, for capturing rare events, e.g. switching between states in a bistable gene circuit, with a switching time of $1000$ min (which give a spectral gap of $\delta=10^{-3}$ events per min), the mixing time for classical Markov chain is $\mathcal{O}(1/\delta)=\mathcal{O}(1000~\text{min})$ per switching event. To observe 100 such switching events, one would need about 70 days ($10^5$ min) of simulation time. Whereas, using quantum walk-based approaches, one can reduce the mixing time to $\mathcal{O}(1/\sqrt{\delta})=\mathcal{O}(32~\text{min})$ and the simulation time for 100 events to 2.2 days.

\section{Whole Cell Spatial Modeling: sample scaling analysis}\label{sec:SI-wcm-spatial}

\subsection{Overview of the WC modeling approach}\label{subsec:wcm-spatial_overview}

Depending on the model scale---spatial or temporal---we focus on, modeling of a whole cell can be broadly described as depicted in \fig\ref{fig:outline}. Johnson et al. \cite{johnson2021quantifying} provides an overview of the various numerical approaches for these systems. On the spatial scale, one can choose to model from the molecular scale all the way to cell division and community interactions. The former requires numerical methods such as molecular dynamics simulations, tracking the system evolution in time scales of $10^{-15}$ to $10^{-7}$ sec, but are intractable at the whole cell scale. Capturing the cell division dynamics, occurring in a time scale of $10^{-6}$ to $10^3$ sec (an $\mathcal{O}(10^{9})$ difference), can be achieved through non-spatial (focusing only on the change in concentration of species – metabolites and proteins) and spatial (focusing on the movement of species within the whole cell) approaches. 

Classical numerical modeling has revolved around deterministic and stochastic approaches for both the non-spatial and spatial approaches. Deterministic approaches involve using deterministic reaction rate equations to model the evolution of the species (concentration and/or spatial dynamics), assuming high concentration and homogeneous spatial distribution of molecular components. Whereas, stochastic approaches capture the more realistic behavior of the individual components by capturing stochastic processes such as thermal Brownian motion and the collisions and interactions among individual particles. Computational complexity of deterministic approaches will be the main focus of the current work as the scale of such problems for small cell systems are tractable using current HPC systems. A whole-cell scale simulation of stochastic processes can be computationally expensive and intractable using the current state-of-the-art high-performance computing (HPC) systems, and for quantum systems would need reformulating the existing governing equations to ``fit" a quantum mechanical model. We discuss a few potential directions for such efforts in the Discussion section (\ref{sec:discussion}).

We broadly describe the deterministic approaches in the following discussion, and propose candidate quantum algorithms that could substitute these classical numerical techniques. We focus on comparing the computational cost for these classical and quantum approaches to perform a single cell cycle of the Yeast cell -- an estimate time of 2 hrs.

\subsection{Non-spatial models}\label{subsec:wcm-spatial_nonspatial}

Deterministic non-spatial models comprise of reaction rate equations, which are coupled ordinary differential equations (ODEs) that determine the change in concentration of various species over time. The procedure to solve such ODEs involves solving a system of linear equations of size $N$ at every time step, where $N$ represents the number of species.

Classical numerical ODE solvers can be explicit or implicit. While explicit solvers have linear computational scaling with respect to the system size, $\mathcal{O}(N)$ for each time step, they have strict stability constraints -- choice of the numerical time-step of the solver is restricted, usually described by the Courant--Friedrichs--Lewy condition of the problem. Whereas, implicit solvers are unconditionally stable, offering the choice of larger time steps, but with the caveat of higher computational cost per time step -- $\mathcal{O}(N^m)$ with $m\ge1$ for non-spatial system with tridiagonal solve \cite{coppersmith1982asymptotic,trefethen2022numerical}. Given that the Jacobian matrix of the reaction rate equations of a cell model need not form a sparse, tridiagonal form, realistically, the computational complexity can be between $m\in[1.5,3]$. Since choosing an explicit solver could potentially lead to an extremely small time step for modeling a cell cycle---$\mathcal{O}(10^9)$ difference between the largest and smallest time scales---we focus our discussion on implicit solvers.

Quantum linear systems algorithms (QLSA) are a class of quantum algorithms that can solve a system of linear equations, which we consider to perform the matrix solve in the implicit solver. The Harrow--Hassidim--Lloyd (HHL) \cite{harrow2009quantum} algorithm is one of the earliest such algorithms that have shown theoretical proofs of attaining quantum advantage. The HHL algorithm delivers a computational complexity of $\mathcal{O}(\log N \kappa^2 s^2 \epsilon^{-1})$, with additional dependence of the condition number of the problem $\kappa$, sparsity of the system matrix $s$, and order of approximation $\epsilon$. There has been various extension of the HHL solver to address the dependence on the condition number \cite{clader2013preconditioned,golden2022quantum,tsemo2024enhancing}, different QLSA algorithms focusing on adiabatic approaches \cite{subacsi2019quantum,an2022quantum,lin2020optimal}, and dependence on the quantum phase estimation algorithm \cite{childs2017quantum,chakraborty2018power,lu2025lugo}, improving the scalability of the approach to a computational complexity of $\mathcal{O}(\log N \kappa \log (\kappa/\epsilon) s^2 )$ \cite{costa2022optimal,jennings2023efficient}. The reader is directed to the review article by \cite{morales2024quantum} for an elaborate survey of QLSA. While we consider using the QLSA as the quantum implicit solver, the practical implementation bottlenecks of data input and output between time-steps is a crucial consideration, which we elaborate in the Discussion section~\ref{sec:discussion}.

\begin{figure}[!ht]
\centering
\includegraphics[width=0.9\textwidth]{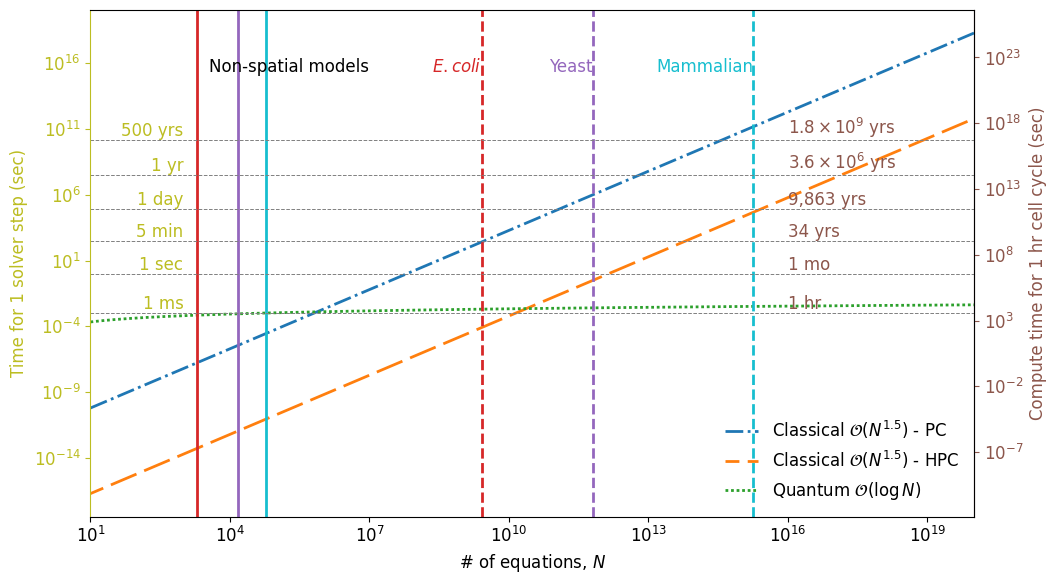}
\caption{\textbf{Classical and quantum linear solver time complexity study.} Time complexity of different implicit solvers for evolution of 1 timestep in a cell simulation (left axis) and for 1~hr of cell cycle (right axis) with increasing system size (number of equations solved). For the classical solution, time taken by the solver on an HPC system (1.6~ExaFLOPS) and a personal computer (500~GigaFLOPS) are shown. The quantum system is assumed to have a throughput of 10.5~kop/s of 16-bit floating point precision \cite{hoefler2023disentangling}. The spatial models assume voxels that are $10 nm$ on edge. The left and right vertical axes denote the wall clock time to solve 1 time step and 1 hr of cell cycle of the model simulation, respectively. Reference vertical lines represent the system sizes for \textit{E. coli}, Yeast, and Mammalian cells.}
\label{fig:time_complexity_spatialwcm}
\end{figure}

To compare the computational cost for the classical and quantum approaches, we consider proteins and metabolites as the species, which for the Yeast cell is around 15,000 \cite{milo2015cell,milo2010bionumbers}. Thus, we are solving for 15,000 coupled ODEs. For the implicit solver, we choose a time step of $1 \times 10^{-3}$ sec to simulate 2 hrs of the Yeast cell cycle. For computing the computational complexity, we choose the best-case scenario for the classical solver---$\mathcal{O}(N^{1.5})$---and for the quantum solver, as a preliminary step of comparison, we choose $\kappa = 1$ and $s = 1$. Figure~\ref{fig:time_complexity_spatialwcm} depicts the complexity of the classical and quantum implicit solvers with respect to the system size - number of equations solved. It is clear that the computational advantage of the quantum implicit solver cannot be attained for solving the non-spatial models, even for a Mammalian cell whose system size is assumed to be $60,000$. Using the above orders of complexity and solving 7.2M time steps for the Yeast cell, we estimate a total time of $8 \mu$sec and $26$ sec for the classical solver using HPC and PC, respectively, whereas $110$ min for the quantum solver.

\subsection{Spatial deterministic models}\label{subsec:wcm-spatial_spatial-deterministic}

To capture a more realistic behavior of a whole cell, in addition to modeling the concentration change of species, deterministic spatial models involve capturing the dynamic movement of the species within the cell. Modeling both the spatial and temporal changes in the position and concentration of the species requires solving coupled partial differential equations (PDEs), which are extensions of the reaction rate equations incorporating diffusion - the reaction-diffusion equations. Spatial models, in addition to capturing the fundamental reaction types in the spatial dimension, need to treat reaction for the molecular exchange between the bulk and the cell surfaces, must specify an equation of motion (usually through diffusion dynamics, but can involve additional features), and boundary conditions of the cell geometry.

Classically, the general procedure involves first spatially discretizing the cell domain as grid points (or cells or elements or voxels). We assume a three-dimensional (3D) spatial model of the cell systems, which is discretized by 3D subvolumes or voxels. Assuming the meshing process is done as a pre-processing step, which can be a computationally intensive procedure to accurately capture the cell boundaries, here, we focus only on the computational complexity of the PDE solver. The next step involves formulating the PDEs as coupled ODEs at the voxels using finite difference (or element or volume) methods. Formulating the PDEs into coupled ODEs can be crucial to form sparse and well-condition systems. Finally, these coupled ODEs are solved at each voxel at each time step using an implicit (or explicit) solver, similar to the non-spatial approach, but with more complex and larger number of equations. The computational cost of 3D implicit solvers for each time step is between $\mathcal{O}(N^{1.5})-\mathcal{O}(N^{2.3})$ based on the sparsity of the system for a direct solve. If one follows an iterative solve, the complexity is $\mathcal{O}(Nk)$ for $k$ iterations, which depends on the condition number (or preconditioning) of the system \cite{coppersmith1982asymptotic,trefethen2022numerical}. 

Similar to the non-spatial approach, QLSA can be used to solve the reaction-diffusion PDEs using a quantum computer at each time step once the system Jacobian is formulated by converting the PDEs to coupled ODEs. While the sparsity of the system can be favourable, the main considerations here for the computational complexity is the impending increase in the condition number of the system due to stiff equations for the cell models. Preconditioning \cite{golden2022quantum} or precondition-free \cite{tsemo2024enhancing} HHL approaches can be useful in such scenarios. For spatial models, while the system size can reach to that needed to demonstrated the quantum advantage over the classical implicit solver (as we will see in Fig.~\ref{fig:time_complexity_spatialwcm}), the main practical bottleneck will be both the data input to the quantum circuits as well as the measurements made from the solution state. We elaborate on considerations to address such bottlenecks in the Discussion section~\ref{sec:discussion}.

\begin{figure}[ht!]
\centering
\includegraphics[width=1\textwidth]{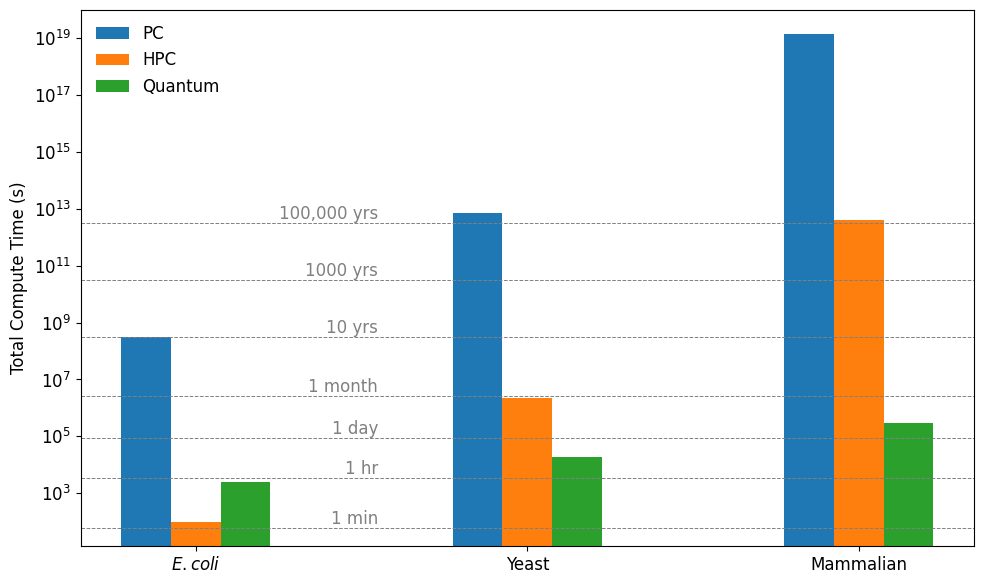}
\caption{\textbf{Computational time for 1 cell cycle of various system.} Total time take by classical and quantum implicit solvers for 1 cell cycle evolution of different whole cell systems -- \textit{E. coli}: 20 min, Yeast: 2 hrs, and Mammalian: 24 hrs.}
\label{fig:total_time}
\end{figure}

For the computational cost comparison, we assume the Yeast cell to have a cell volume of $42\,\mu\mathrm{m}^{3}$, which is discretized using voxels with $10\,\mathrm{nm}$ edges so as to capture at least one protein ($4-5\,\mathrm{nm}$ in length) \cite{milo2015cell,milo2010bionumbers} and several small molecule metabolites. Thus, for 15,000 species, we need to solve $6.3\times 10^{11}$ coupled ODEs. We use the same computational complexity for the classical and quantum implicit solvers as chosen for evaluating the non-spatial model. Figure~\ref{fig:time_complexity_spatialwcm} depicts the computational complexity of the solvers with increase in system size, with reference system sizes of \textit{E. coli}, Yeast, and Mammalian cells. The results suggest that the quantum solver can potentially demonstrate advantage over the classical approach for simulating the Yeast cell. Choosing a time step of $1 \times 10^{-3}$ sec to simulate 2 hrs of the Yeast cell cycle, we estimate a total time of $7.2\times10^{12}$ sec (over 228,000 years) and $2.25\times10^6 $ sec ($\approx26$ days) for the classical solver using the PC and HPC, respectively, whereas only $1.8\times10^4$ sec (5 hrs) for the quantum solver. The total time taken by the solvers for various cell systems are shown in \fig\ref{fig:total_time}.

\subsection{Spatial stochastic approach}\label{subsec:wcm-spatial_spatial-stochastic}

While the deterministic approach assumes a homogeneous distribution of species concentrations within a cell, capturing the discrete and stochastic nature of biochemical systems requires a stochastic spatial approach.  This is achieved through the Reaction--Diffusion Master Equation (RDME), which couples the CME within each spatial voxel to diffusion jumps between neighboring voxels. In the RDME framework, each voxel runs its own stochastic simulation, giving rise to an effective number of reaction channels per voxel that includes both local biochemical reactions and diffusion events across the six voxel faces:
\begin{equation}
  R_{\mathrm{total}} = R_r + 6\,M,
  \label{eq:R_total}
\end{equation}
where $M$ is the number of molecular species and the factor of six accounts for diffusion across each face of a cubic voxel.

For the computational cost comparison, we adopt the same spatial discretizations of the \textit{E.~coli} cell as in SI~\ref{subsec:wcm-spatial_spatial-deterministic}: a cell volume of $1\,\mu\mathrm{m}^{3}$ discretized into voxels with $10\,\mathrm{nm}$ edges, yielding $N_{\mathrm{voxels}} = 10^{6}$ voxels. With $M = 4000$ species for \textit{E.~coli} and $R_r = 30$ reaction channels (glycolysis pathway, SI~\ref{subsec:SI-wcm-network_stochastic}), the effective number of channels per voxel is $R_{\mathrm{total}} \approx 2.4 \times 10^{4}$ channels/voxel. The total propensity for the glycolysis pathway in \textit{E.~coli} is $a_{0} = 10^{9}$~events/sec for the whole cell (SI~\ref{subsec:SI-wcm-network_stochastic}). Normalized across all voxels, the per-voxel propensity is
\begin{equation}
  a_{0}^{\mathrm{voxel}}
    = \frac{10^{9}}{10^{6}}
    = 10^{3}\ \text{events/sec/voxel}.
\end{equation}

For one \textit{E.~coli} cell cycle ($T_{\mathrm{sim}} = 20\,\mathrm{min} =
1200\,\mathrm{sec}$), the number of reaction events per voxel is
\begin{equation}
  N_{r}^{\mathrm{voxel}}
    = a_{0}^{\mathrm{voxel}} \times T_{\mathrm{sim}}
    = 1.2 \times 10^{6}\ \text{events/voxel}.
\end{equation}

The total computational cost for a single spatial SSA trajectory is therefore
\begin{equation}
  N_{\mathrm{total}}
    = R_{\mathrm{total}} \times N_{r}^{\mathrm{voxel}} \times N_{\mathrm{voxels}}
    = 2.4 \times 10^{4}
      \times 1.2 \times 10^{6}
      \times 10^{6}
    = 2.9 \times 10^{16}\ \text{operations}.
\end{equation}

\paragraph{Classical (single core):}
Using an optimized single-core C++ SSA implementation on a modern workstation CPU (${\sim}10^{7}$~events/sec), the wall-clock time for a single trajectory is
\begin{equation}
  T_{\mathrm{serial}}
    = \frac{2.9 \times 10^{16}}{10^{7}}
    \approx 2.9 \times 10^{9}\ \mathrm{sec}
    \approx 92\ \text{years}.
\end{equation}

\paragraph{Classical (HPC -- OLCF Frontier):}
Unlike the non-spatial SSA case (SI~\ref{subsec:SI-wcm-network_stochastic}), the RDME admits spatial domain decomposition: distinct voxel subdomains can be assigned to different CPU cores, with inter-voxel diffusion jumps handled as boundary communication.  Frontier provides $9{,}856 $ nodes each with $64$ CPU cores, giving $N_{\mathrm{cores}} \approx 630{,}784$ cores in total and a favorable load balance of $\approx 2$ voxels/core. Assuming ideal strong scaling (no communication overhead), the wall-clock time is
\begin{equation}
  T_{\mathrm{Frontier}}^{\mathrm{ideal}}
    = \frac{2.9 \times 10^{16}}{630{,}784 \times 10^{7}}
    \approx 4.6 \times 10^{3}\ \mathrm{sec}
    \approx 1.3\ \text{hrs}.
\end{equation}
However, fine-grained spatial decomposition of the RDME incurs significant
inter-voxel communication overhead at voxel boundaries (typically a $10$--$100
\times$ penalty for nearest-neighbor diffusion exchange), yielding a realistic
wall-clock estimate of approximately
\begin{equation}
  T_{\mathrm{Frontier}}^{\mathrm{realistic}}
    \approx 13\ \text{hrs}\ \text{to}\ 5.5\ \text{days}.
\end{equation}

\paragraph{Quantum.}
Quantum Markov chain and quantum walk-based algorithms~\cite{szegedy2004quantum,layden2023quantum} can target the RDME operator directly, encoding the effective state-space of the system as a linear operator of dimension
\begin{equation}
  N
    = R_{\mathrm{total}} \times N_{\mathrm{voxels}}
    = 2.4 \times 10^{4} \times 10^{6}
    = 2.4 \times 10^{10},
\end{equation}
giving $\log N \approx 11$ operations/trajectory using a QLSA. Note that $N_{r}^{\mathrm{voxel}}$ does not appear here because the QLSA solves the linear system encoding the RDME generator in a single operator application, rather than stepping through each reaction event sequentially as in classical SSA. At the assumed quantum hardware throughput of $10.5$~kop/s (SI~\ref{subsec:wcm-spatial_overview}), and
requiring $\mathcal{O}(1/\varepsilon) \approx 20$ quantum samples for
$5\%$ target error via amplitude estimation~\cite{an2021quantum}, the total simulation
time is
\begin{equation}
  T_{\mathrm{quantum}}
    = 20 \times \frac{11}{10500}
    \approx 0.02\ \mathrm{sec}.
\end{equation}

The computational costs for the stochastic spatial model of one \textit{E.~coli}
cell cycle are summarised in Table~\ref{tab:D4_comparison}.

\begin{table}[h]
\centering
\caption{Computational time comparison for a stochastic spatial (RDME) simulation
  of one \textit{E.~coli} cell cycle ($T_{\mathrm{sim}} = 20$~min,
  $10^{6}$ voxels, $4{,}000$ species, $R_{\mathrm{total}} \approx 2.4 \times
  10^{4}$ channels/voxel). Quantum complexities assume oracle access, $s=1$,
  $\kappa=1$, and omit data-loading costs.}
\label{tab:D4_comparison}
\begin{tabular}{llll}
\hline
\textbf{Implementation} & \textbf{Wall-clock time} & \textbf{Notes} \\
\hline
Single-core C++ (workstation) & $\sim$92 years  & Serial; $10^{7}$~events/sec \\
Frontier (idealized)          & $\sim$1.3 hrs   & Parallel through domain decomposition; no communication overhead \\
Frontier (realistic)          & $\sim$13 hrs -- 5.5 days & With inter-voxel communication overhead \\
Quantum (QLSA + amp.\ est.)   & $\sim$0.02 sec   & $\log N \approx 11$; idealized $\kappa$, $s$ \\
\hline
\end{tabular}
\end{table}

\end{document}